\documentclass[12pt,a4paper]{article}
\usepackage{amsmath}
\usepackage[dvips]{graphicx}
\input epsf
\usepackage{times}
\begin{document}
\begin{titlepage}
\title{Unitarity and geometrical aspects at low $x$}
\author{{S. M. Troshin}%\footnote{e-mail: troshin@mx.ihep.su\rm}
, N. E. Tyurin\\[1ex]
\small  \it Institute for High Energy Physics,\\
\small  \it Protvino, Moscow Region, 142280 Russia}
\normalsize
\date{}
\maketitle

\begin{abstract}
On the grounds of  extension of the $U$--matrix unitarization to
the off--shell scattering we consider virtual photon induced
scattering. We discuss  behaviour of the structure function
$F_2(x,Q^2)$ at low $x$ and the total cross--section of virtual
photon--proton scattering and obtain, in particular, the dependence
$\sigma^{tot}_{\gamma^* p}\sim (W^2)^{\lambda(Q^2)}$ where
exponent   $\lambda(Q^2)$  is related to the interaction radius of
a constituent quark.\\[2ex]
\end{abstract}
\end{titlepage}
\setcounter{page}{2}

\section*{Introduction}

Experimental data obtained at HERA  \cite{her}  clearly
demonstrated rising behaviour of the structure function
$F_2(x,Q^2)$ at small $x$ which is translated to the rising
dependence of
 the  total cross--section
 $\sigma^{tot}_{\gamma^*p}(W^2, Q^2)$
 on center of mass energy $W^2$.
This effect
 is consistent with various dependencies
on $W^2$ and has been treated in different ways, e. g.  as a manifestation of
 hard BFKL Pomeron \cite{lipa}, a confirmation of the DGLAP evolution
 in the perturbative QCD \cite{pqcd},
 a transient phenomena, i.e. preasymptotic effects \cite{nad}
 or as a true asymptotical
dependence of the off--mass--shell scattering amplitude
\cite{petr}. This list is far from being complete and other
interpretations can be found, e.g. in the review papers \cite{her,lands}.

It is worth to  note here that
the  essential point in the study of low-$x$ dynamics
is that the space-time structure of
the scattering
at small values of $x$ involves  large distances
$l\sim 1/mx$ on the light--cone \cite{pas} and the
region $x\sim 0$ is  sensitive to the  nonperturbative contributions.
The deep--inelastic scattering in this region turns out to be a coherent
process where diffraction plays a major role and nonperturbative models
such as  Regge or vector dominance model can be competitive with perturbative
QCD successfully applied for  description of the experimental data.

The
strong experimentally observed rise of
$\sigma^{tot}_{\gamma^*p}(W^2,Q^2)$ when
\begin{equation}\label{wlq}
\sigma^{tot}_{\gamma^*p}(W^2,Q^2)\propto (W^2)^{\lambda(Q^2)}
\end{equation}
with $\lambda(Q^2)$ rising with $Q^2$ from about $0.1$ to about
$0.4$
 was considered to a somewhat extent as a surprising fact
on the grounds of our knowledge of the
 energy dependence of total cross--sections in hadronic
 interactions, where $\lambda\sim 0.1$.
The above comparison between photon--induced and hadron--induced
interactions is quite legitimate since the photon is demonstrating
its hadronlike nature for a long time.  The apparent difference
between the hadron and virtual photon total cross--section
behaviours however has no fundamental meaning since there is no
Froissart--Martin bound in the case
off--shell particles \cite{petr,indur}. Only under some additional
assumptions this bound can be applied \cite{ttpre,levin}.

This
 problem was addressed in \cite{prokud} on the basis of
unitarity for off--shell scattering starting from the eikonal
representation for the scattering amplitude. It was argued that
the observed rise of
 $\sigma^{tot}_{\gamma^*p}(W^2,Q^2)$  can be considered
as a  true asymptotic behaviour and extension of the eikonal
representation for off-shell particles does not provide
limitations for  $\sigma^{tot}_{\gamma^*p}(W^2,Q^2)$ at large $W^2$.
It was claimed that HERA data can be described by the classical
universal (with $Q^2$--independent intercepts) Regge trajectories.

In the present paper we treat similar problems on the basis of the
off--shell $U$--matrix approach to the amplitude
unitarization.  It is
shown that the unitary representation for  off--shell particles
and the respective extension of the chiral quark model for the
 $U$--matrix can lead to  Eq. (\ref{wlq}), where
the exponent $\lambda (Q^2)$ is related to the $Q^2$--dependent
 interaction radius of a virtual (off--shell) constituent quark.

It is to be stressed here the importance of the effective interaction
radius concept \cite{log}.  The study of the effective interaction radius
dependence on the scattering variables seemed very useful for
understanding of the relevant dynamics of high energy hadronic
reactions \cite{chy,khru}.
Now on it is widely known that the respective
geometrical considerations about interaction provide a deep insight
in hadron dynamics and deep--inelastic scattering
(cf.  \cite{bartels}).
\section{Off--shell  scattering  in the $U$--matrix method}
The extension of the $U$--matrix unitarization for the off-shell
scattering was considered in \cite{ttpre}. It was supposed that
the virtual  photon fluctuates into the virtual  vector
meson states which afterwards interact with a hadron. We
considered  a single effective vector meson field and used the
notions $F^{**}(s,t,Q^2)$, $F^{*}(s,t,Q^2)$ and $F(s,t)$ for the
amplitudes when both initial and final mesons  are off mass
shell, only initial meson is off mass shell and both mesons are on
mass shell, correspondingly.  The virtualities of initial and
final vector mesons were chosen to be equal $Q^2$ since we  need  the
amplitude of the forward virtual Compton scattering.

The equations for the amplitudes $F^{**}$ and $F^*$ have the same
structure as the equation for the on--shell amplitude $F$ but
relate the different amplitudes.
 In the impact
parameter representation ($s\gg 4m^2$) they can be written as follows
\begin{eqnarray}
F^{**}(s,b,Q^2) & = & U^{**}(s,b,Q^2)+iU^{*}(s,b,Q^2)F^{*}(s,b,Q^2)\nonumber\\
F^{*}(s,b,Q^2) & = & U^{*}(s,b,Q^2)+iU^{*}(s,b,Q^2)F^{}(s,b).\label{es}
\end{eqnarray}
The solutions  are
\begin{eqnarray}
F^*(s,b,Q^2) & = & \frac{U^*(s,b,Q^2)}{1-iU(s,b)},\label{*}\\
F^{**}(s,b,Q^2) & = & \frac{U^{**}(s,b,Q^2)}{1-iU(s,b)}+
i\frac{[U^{*}(s,b,Q^2)]^2-U^{**}(s,b,Q^2)U(s,b)
}{1-iU(s,b)}.\label{**}
\end{eqnarray}
We also assumed the following relation
\begin{equation}
[U^{*}(s,b,Q^2)]^2-U^{**}(s,b,Q^2)U(s,b)=0.\label{zr}
\end{equation}
which can be identically fulfilled  if the following factorization
occurs:
\begin{eqnarray}
U^{**}(s,b,Q^2) & = & \omega ^2(s,b,Q^2)U(s,b)\nonumber\\
U^{*}(s,b,Q^2) & = & \omega(s,b,Q^2)U(s,b).\label{fct}
\end{eqnarray}
Such factorization is valid, e. g. in the Regge model with
factorizable residues and the $Q^2$--independent trajectory. It is
also valid in the off--shell extension of the chiral quark model
for the $U$--matrix. We consider the latter further in
detail.

Thus, we  have for the amplitudes $F^*$ and $F^{**}$
\begin{eqnarray}
F^{*}(s,b,Q^2) & = & \frac{U^{*}(s,b,Q^2)}{1-iU(s,b)}=
\omega(s,b,Q^2)\frac{U(s,b)}{1-iU(s,b)}
\label{vrq}\\
F^{**}(s,b,Q^2) & = & \frac{U^{**}(s,b,Q^2)}{1-iU(s,b)}=
\omega^2(s,b,Q^2)\frac{U(s,b)}{1-iU(s,b)} \label{vr}
\end{eqnarray}
and unitarity  provides inequalities
\begin{eqnarray}
|F^*(s,b,Q^2)| & \leq & |\omega (s,b,Q^2)|,\nonumber\\
|F^{**}(s,b,Q^2)| & \leq & |\omega^2(s,b,Q^2)|.\label{bnd}
\end{eqnarray}

To discuss the asymptotical behaviour of
$\sigma^{tot}_{\gamma^* p}$
 we consider  off--shell extension of the model for
hadron scattering \cite{csn}, which is based on the ideas of chiral quark
models.   Valence quarks located in the
central part of a hadron are supposed to scatter in a
quasi-independent way by the effective field.  In accordance with the
quasi-independence of valence quarks we represent the basic
dynamical  quantity in  the form  of  product:
\begin{equation} U(s,b)\,=\, \prod^{n_{h_1}}_{i=1}\, \langle f_{Q_i}(s,b)
\rangle \prod^{n_{h_2}}_{j=1}\, \langle f_{Q_j}(s,b)
\rangle\label{prd} \end{equation} in the impact parameter
representation, $N=n_{h_1}+n_{h_2}$ is the total number of
constituent quarks in the initial hadrons.  Factors $\langle
f_{Q}(s,b)\rangle$ correspond to the individual quark
scattering
 amplitude smeared over transverse position of $Q$ inside hadron
  $h$ and over fraction of longitudinal momentum of the initial
  hadron carried by quark $Q$.
Factorization Eq. (\ref{prd}) reflects the coherence in the valence
quark scattering and may be considered as an effective
implementation of constituent quarks' confinement. This mechanism
resembles  Landshoff mechanism of quark--quark independent
  scattering \cite{landso}.  However, in this case we refer
   not to pair
 interaction of valence quarks belonging to the colliding hadrons, but
rather to Hartree--Fock approximation for the
constituent quark scattering in the mean field.

 The  picture of hadron structure  implies  that     the
overlapping of hadron structures and interaction of the
condensates
 occur  at  the first stage of collision. Due to excitation of the condensates,
the quasiparticles, i.e. massive quarks arise. These quarks play role
of scatterers.
 To estimate number
of such quarks one could assume that  part of hadron energy carried by
the outer condensate clouds is being released in the overlap region
 to
generate massive quarks. Then their number can be estimated  by
the quantity:
 \begin{equation} \tilde{N}(s,b)\,\propto
\,\frac{(1-k_Q)\sqrt{s}}{m_Q}\;D^{h_1}_c\otimes D^{h_2}_c,
\label{4}
\end{equation} where $m_Q$ -- constituent quark mass, $k_Q$ -- hadron
 energy fraction
  carried  by  the constituent valence quarks. Function $D^h_c$
describes condensate distribution inside the hadron $h$, and $b$ is
an impact parameter of the colliding hadrons $h_1$ and $h_2$.
Thus, $\tilde{N}(s,b)$ quarks appear in addition to $N$
valence quarks. Those quarks are transient
ones: they are transformed back into the condensates of the final
hadrons in elastic scattering. It should be noted that we use subscript
$Q$ to refer  the  constituent quark $Q$ and the same letter $Q$
is used to denote a virtuality $Q^2$. However, they enter formulas
in a way excluding  confusion.

 The amplitude $\langle f_Q(s,b)\rangle $ describes elastic
   scattering $Q\to Q$ of a single
valence  on-shell quark $Q$  in the effective field and
 we use for the function $\langle
f_Q(s,b)\rangle $  the following expression
\begin{equation}
\langle f_Q(s,b)\rangle =[\tilde{N}(s,b)+(N-1)] \,V_Q(\,b\,),
\label{7}
\end{equation}
where $V_Q(b)$  has a simple form
 \begin{equation}
 V_Q(b)\propto g\exp(-m_Qb/\xi ),
 \end{equation}
which corresponds to the quark interaction radius
 \begin{equation}
r_Q=\xi/m_Q.
\end{equation}
This picture can be extended for the case when the hadron $h_2$
(vector meson in our case)  is off mass shell.  The off--shell
$U$--matrix, i.e.  $U^{**}(s,b,Q^2)$ should be then presented as the
 product
\begin{equation} U^{**}(s,b,Q^2)\,=\, \prod^{n_{h_1}}_{i=1}\,
\langle f_{Q_i}(s,b) \rangle \prod^{n_{h^*_2}}_{j=1}\, \langle
f^{}_{Q^*_j}(s,b,Q^2) \rangle\label{prdv} \end{equation} and
the function $\langle f_Q^*(s,b,Q^2)\rangle $ is to be written
as
\begin{equation}
\langle f^{}_{Q^*}(s,b,Q^2)\rangle =[\tilde{N}(s,b)+(N-1)] \,V_{Q^*}(b, Q^2).
\label{fqv}
\end{equation}
The notion $\langle f^{}_{Q^*}(s,b,Q^2)\rangle$ stands for
the the smeared amplitude which describes elastic
scattering $Q^*\to Q^*$ of a single
valence constituent off-shell quark $Q^*$  in the effective field.
In the above equation
\begin{equation}
V_{Q^*}(b,Q^2)\propto g(Q^2)\exp(-m_Qb/\xi (Q^2) )
\end{equation}
and it corresponds to the virtual constituent quark
interaction radius
\begin{equation}\label{rqvi}
r_{Q^*}=\xi(Q^2)/m_Q.
\end{equation}
Under this we mean constituent quark composing effective virtual
 meson.

The $b$--dependence of $\tilde{N}(s,b)$ is weak compared to the
$b$--dependence of $V_Q$ \cite{ttore} and therefore we have taken this function
to be independent on the impact parameter $b$.

Dependence on virtuality $Q^2$ comes through
dependence of the intensity of the virtual constituent
 quark interaction $g(Q^2)$
and the parameter $\xi(Q^2)$, which determines the quark
interaction radius (in the on-shell limit $g(Q^2)\to g$ and
$\xi(Q^2)\to\xi$).

Following these considerations, the explicit dependencies
of the generalized
reaction matrices $U^*$ and $U^{**}$ on $s$, $b$ and $Q^2$
 can easily be written in the form of Eq. (\ref{fct}) with
\begin{equation}\label{omeg}
  \omega(s,b,Q^2)=\frac{\langle f^{}_{Q^*}(s,b,Q^2)\rangle}
  {\langle f^{}_{Q}(s,b)\rangle}.
\end{equation}
Note that Eqs. (\ref{zr}) and (\ref{fct}) imply that
\[
\langle f^{}_{Q^*\to Q}(s,b,Q^2)\rangle=
[\langle f^{}_{Q^*}(s,b,Q^2)\rangle \langle f^{}_{Q}(s,b)\rangle]^{1/2}.
\]

We consider high--energy limit and for simplicity suppose
that all constituent quarks have equal masses and parameters
$g$ and $\xi$ as well as $g(Q^2)$ and $\xi(Q^2)$
do not depend on quark flavor. We also consider for simplicity
pure imaginary amplitudes. Then we have for the functions
$U$, $U^*$ and $U^{**}$ the following expressions
\begin{eqnarray}
 & & U(s,b)  =  ig^N\left (\frac{s}{m^2_Q}\right )^{N/2}
\exp \left [-\frac{m_QNb}{\xi}\right ] \label{usb}\\[1ex]
& & U^*(s,b,Q^2)  =  \omega (b,Q^2) U(s,b)\label{uv}\\[1ex]
& & U^{**}(s,b,Q^2)  =  \omega^2 (b,Q^2) U(s,b)\label{uvv}
\end{eqnarray}
where the function $\omega$ is an energy-independent one
and has the following dependence on $b$ and $Q^2$
\begin{equation}\label{ome}
  \omega(b,Q^2)  =
  \frac{g(Q^2)}{g}\exp \left [-\frac{m_Qb}{\bar{\xi}(Q^2)}\right ]
\end{equation}
with
\begin{equation}\label{ksi}
  \bar{\xi}(Q^2)=\frac{\xi\xi(Q^2)}{\xi-\xi(Q^2)}.
\end{equation}
It is clear that for
 the on--shell particles $\omega \to 1$ and using
  Eqs. (\ref{*}) and (\ref{**})
we will arrive at large $W^2$ to the result obtained in \cite{ttpre}
\begin{equation}\label{ons}
  \sigma^{tot}_{\gamma p}(W^2)\propto\frac{\xi^2}{m_Q^2}\ln ^2 \frac{W^2}{m_Q^2},
\end{equation}
where the usual for deep--inelastic scattering notation $W^2$
 instead of $s$ is used. Similar result is valid also for the off mass shell
 particles when the interaction radius of virtual quark does not depend
 on $Q^2$ and is equal to the interaction radius of the on--shell quark,
 i.e. $\xi(Q^2)\equiv \xi $. The behaviour of the total cross--section
 at large $W^2$
\begin{equation}\label{ofs}
  \sigma^{tot}_{\gamma^* p}(W^2)\propto
  \left[\frac{g(Q^2)\xi}{gm_Q}\right]^2
  \ln ^2 \frac{W^2}{m_Q^2},
\end{equation}
 corresponds to the result obtained in \cite{ttpre}.

For the off--shell case (and $\xi(Q^2)>\xi$) the situation is different
and we have
 at large
$W^2$
\begin{equation}\label{totv}
\sigma^{tot}_{\gamma^* p}(W^2,Q^2)\propto G(Q^2)\left(\frac{W^2}{m_Q^2}
\right)^{\lambda (Q^2)}
\ln \frac{W^2}{m_Q^2},
\end{equation}
where
\begin{equation}\label{lamb}
\lambda(Q^2)=\frac{\xi(Q^2)-\xi}{\xi(Q^2)}.
\end{equation}
We shall further concentrate  on this as we currently
think the most interesting case.

However, it should be noted that for $\xi(Q^2)<\xi$
\[
\sigma^{tot}_{\gamma^* p}(W^2)\propto\left[\frac{g(Q^2)\xi}
{g\lambda(Q^2)m_Q}\right]^2,
\]
i. e. asymptotically cross--section would be energy-independent and this
option cannot be excluded in principle,
 since we are dealing with the limit $W^2\to\infty$.
The last
 scenario would  mean that the experimentally observed rise of
  $\sigma^{tot}_{\gamma^* p}$
 is transient preasymptotic phenomena \cite{nad,ttpre}.
  It can be realized when we
 replace in the formula for the interaction radius of the on--shell constituent
  quark $r_Q=\xi/m_Q$ the mass $m_Q$ by
   $m_{Q^*}=\sqrt{m_Q^2+Q^2}$ in order to obtain
  the interaction radius of the off-shell constituent quark and
  write it down as $r_{Q^*}=\xi/m_{Q^*}$, or equivalently
  replace  $\xi(Q^2)$ for
$\xi(Q^2)={\xi m_Q}/{\sqrt{m_Q^2+Q^2}}$.

All the above expressions for $ \sigma^{tot}_{\gamma^* p}(W^2)$ can
be rewritten as the corresponding dependencies of $F_2(x,Q^2)$ at small $x$
according to the relation $F_2(x,Q^2)=\frac{Q^2}{4\pi^2\alpha}
 \sigma^{tot}_{\gamma^* p}(W^2)$ where $x=Q^2/W^2$. In particular, the
 Eq. (\ref{totv}) will appear in the  form
\begin{equation}\label{totv1}
F_2(x,Q^2)\propto\tilde{G}(Q^2)\left(\frac{1}{x}\right)^{\lambda (Q^2)}
\ln (1/x),
\end{equation}

\section{Phenomenological implications}
It is interesting that the value and $Q^2$ dependence of the
 exponent $\lambda(Q^2)$ is related to the interaction radius
 of the virtual constituent quark. The value of parameter $\xi$
 in the model is determined by the slope of the differential cross--section
of elastic scattering at large $t$ \cite{ttore}, i. e.
\begin{equation}\label{ore}
  \frac{d\sigma}{dt}\propto\exp\left[-\frac{2\pi\xi}{m_QN}\sqrt{-t}\right].
\end{equation}
and from $pp$-experimental data $\xi=2-2.5$. Uncertainty
is related to the ambiguity in the constituent quark mass value. Using for
simplicity $\xi=2$ and the data for $\lambda(Q^2)$ obtained
at HERA \cite{bart} we calculated ``experimental'' $Q^2$--dependence
 of the function
$\xi(Q^2)$:
\begin{equation}\label{ksiq}
\xi(Q^2)=\frac{\xi}{1-\lambda(Q^2)}.
\end{equation}

Results are represented in Figure 1. It is clear that experiment
 leads to $\xi(Q^2)$ rising with $Q^2$. This rise is slow and consistent with
 $\ln Q^2$ extrapolation. Thus, assuming this dependence to be kept
 at higher $Q^2$ and using Eq. (\ref{lamb}), we  predict saturation
 in the $Q^2$--dependence of $\lambda(Q^2)$, i.e. at large $Q^2$ the
 flattening will take place.
\section*{Conclusion}
We considered limitations unitarity provides for the $\gamma^* p$--total
 cross-sections and geometrical effects in the
  model dependence of $\sigma^{tot}_{\gamma^* p}$.
In particular, it was shown that the constituent quark's
 interaction radius  dependence on $Q^2$  can lead to a nontrivial,
asymptotical result: the
behaviour of $\sigma^{tot}_{\gamma^* p}$ will be
$\sigma^{tot}_{\gamma^* p}\sim (W^2)^{\lambda(Q^2)}$,
where $\lambda(Q^2)$ will be saturated at large values of $Q^2$.
This result is valid  when the interaction radius of the virtual
constituent quark is rising with virtuality $Q^2$. The reason
 for such rise
 might be of a dynamical nature and it could originate from
 the emission of the additional $q\bar q$--pairs in the
 nonperturbative  structure of a constituent quark.
In the present
approach constituent quark consists of a current quark
and the  cloud of quark--antiquark pairs of the different
flavors \cite{ttore}.
Available experimental data
 are consistent with the dependence $\xi(Q^2)\propto \ln Q^2$.

We would also like to note that unitarity is the most essential
property the asymptotical behaviour of the cross--section
$\sigma^{tot}_{\gamma^* p}$ depends on, and the above power energy
 dependence is not
 the only possible one. Unitarity transforms strong energy
dependence of the ``Born'' amplitude $\sim (W^2)^{N/2}$ into
$\ln^2 W^2$ dependence for the on--shell particles or when interaction
 radius of the virtual constituent quark does not depend on virtuality
 and equal to the interaction radius of the  on--shell constituent
  quark.
 Unitarity transforms this strong energy behaviour into the
 one, which depends on energy mildly  $\sim(W^2)^{\lambda(Q^2)}$
  when $\xi(Q^2)>\xi$. It
 can even lead to the asymptotically constant cross--section
$\sigma^{tot}_{\gamma^* p}$ when $\xi(Q^2)<\xi$.

  The available experimental data for the structure functions
  at low values of $x$ continue to demonstrate the
  rising total cross-section of $\gamma^* p$--interactions
 and therefore we currently consider rising with virtuality
  interaction radius of a constituent quark
  as a most relevant  option, however, it does not mean that the
  other possibilities are principally excluded.

\section*{Acknowledgements} We would like to thank  V. A. Petrov
for the stimulating discussions. This work was supported in part
 by  RFBR under Grant 99-02-17995.

\newpage
\normalsize

\newpage

\begin{figure}[htb]
\vspace{2mm}
\begin{center}
\epsfxsize=5in \epsfysize=5in
 \epsffile{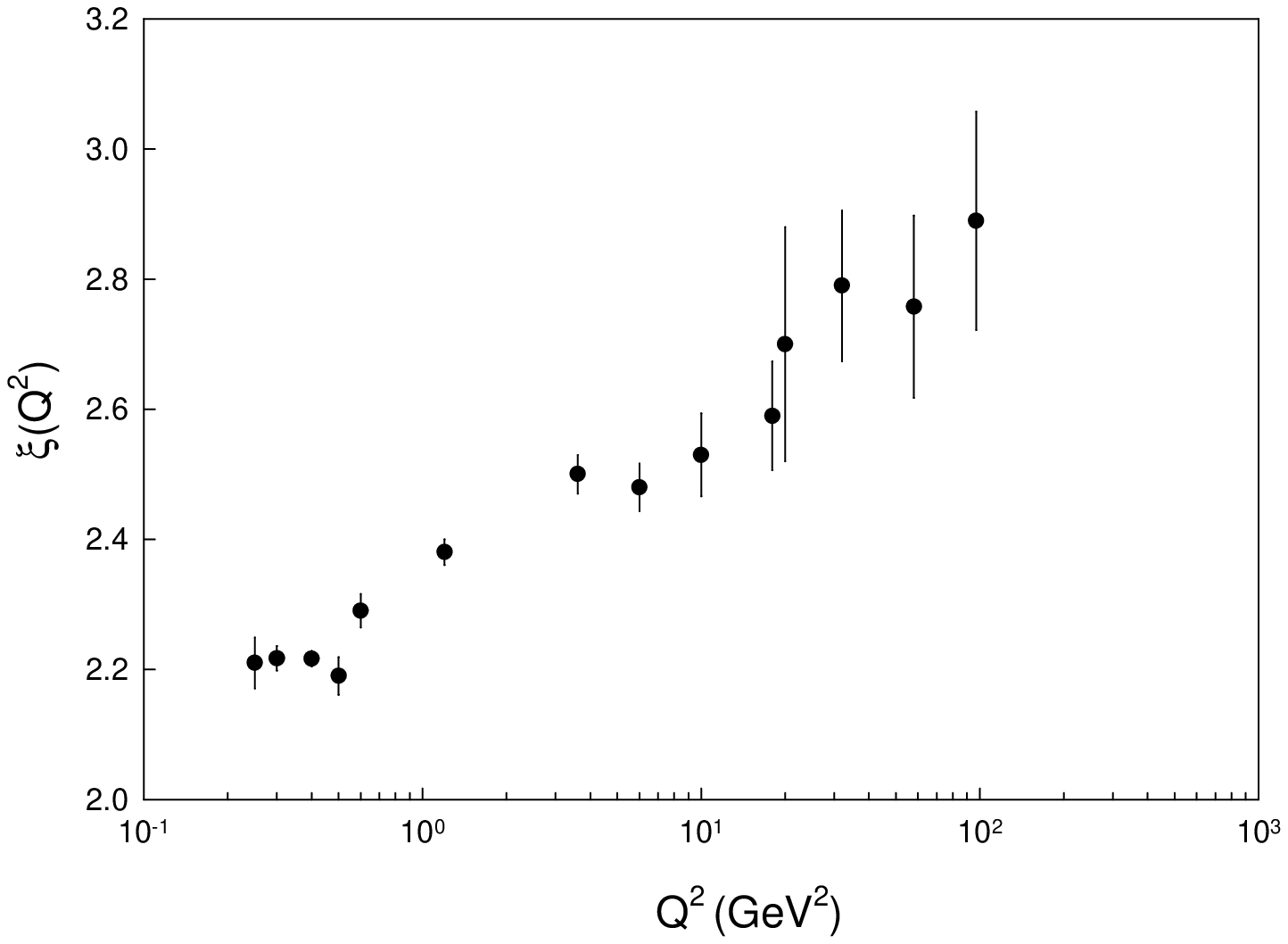}
\end{center}
 \caption[ksi]{ The ``experimental'' data for the function $\xi(Q^2)$.}
\label{fig:1}
\end{figure}

\end{document}